\documentclass[pre,aps,showkeys,showpacs,superscriptaddress,twocolumn,%
amsmath,amssymb]{revtex4} 

\usepackage{hyperref}
\usepackage{graphics}
\usepackage{epsfig}
\usepackage{epstopdf}
\usepackage{amsmath}
\usepackage{setspace}
\usepackage{graphicx}
\usepackage{dcolumn}
\usepackage{bm}
\usepackage{color}
\usepackage{latexsym}
\usepackage{textcomp}

\newcommand{\textmark}[1]{\textcolor{black}{ #1}}

\begin{document}

\title{Timescale divergence at the shear jamming transition}

\author{H. A. Vinutha}
\email{vh327@cam.ac.uk}
\affiliation{Jawaharlal Nehru Center for Advanced Scientific Research,
Jakkur Campus, Bengaluru 560064, India}
\affiliation{Centre for Interdisciplinary Sciences, Tata Institute of Fundamental Research, Hyderabad 500107, India}
\author{Kabir Ramola} 
\email{kramola@tifrh.res.in}
\affiliation{Centre for Interdisciplinary Sciences, Tata Institute of Fundamental Research, Hyderabad 500107, India}
\author{Bulbul Chakraborty}
\email{bulbul@brandeis.edu}
\affiliation{Martin Fisher School of Physics, Brandeis University, Waltham, MA 02454, USA}
\author{Srikanth Sastry}
\email{sastry@jncasr.ac.in}
\affiliation{Jawaharlal Nehru Center for Advanced Scientific Research,
Jakkur Campus, Bengaluru 560064, India}
\affiliation{Centre for Interdisciplinary Sciences, Tata Institute of Fundamental Research, Hyderabad 500107, India}

\date{\today} 


\begin{abstract}
We find that in simulations of quasi-statically sheared frictional disks, the shear jamming transition can be characterized by an abrupt jump in the number of force bearing contacts between particles. This {\it mechanical} coordination number increases discontinuously from $Z = 0$ to $Z \gtrsim d +1$ at a critical shear value $\gamma_c$, as opposed to a smooth increase in the number of geometric contacts.
This is accompanied by a diverging timescale $\tau^*$ that characterizes the time required by the system to attain force balance when subjected to a perturbation. As the global shear $\gamma$ approaches the critical value $\gamma_c$ from below, one observes the divergence of 
the time taken to relax to a state where all the inter-particle contacts have uniformly zero force.  Above $\gamma_{c}$, the system settles into a state characterized by finite forces between particles, with the timescale also increasing as $\gamma \to \gamma_{c}^{+}$. By using two different protocols to generate force balanced configurations, we show that this timescale divergence is a robust feature that accompanies the shear jamming transition. 
\end{abstract} 


\maketitle


\subsection{Introduction}

Rigidity, the ability of solids to sustain finite stresses, arises in crystalline systems due to their broken translational symmetry \cite{chaikin2000principles}. However, a wide variety of disordered materials like glasses, granular packings, suspensions, colloids, and gels exhibit rigidity even when  translational symmetry is not broken. Rigidity in amorphous packings can be induced by the rapid cooling of liquids \cite{debenedetti2001supercooled}, increasing the density of finite-sized particles \cite{o2003jamming} or by the application of a shear deformation \cite{cates1998jamming,bi2011}. An interesting example is a packing of  frictional athermal disks undergoing shear. As the shear is increased, contacts develop between the disks. When a critical number of contacts per particle $Z_c = d + 1$ are created, the system is able to sustain external stresses, a phenomenon termed {\it shear jamming} \cite{bi2011}. Shear thus provides a different control parameter with which to explore the behavior of systems close to jamming, particularly in the context of dense suspensions \cite{wyart2014discontinuous}.

\begin{figure}
\includegraphics[scale=0.65,angle=0]{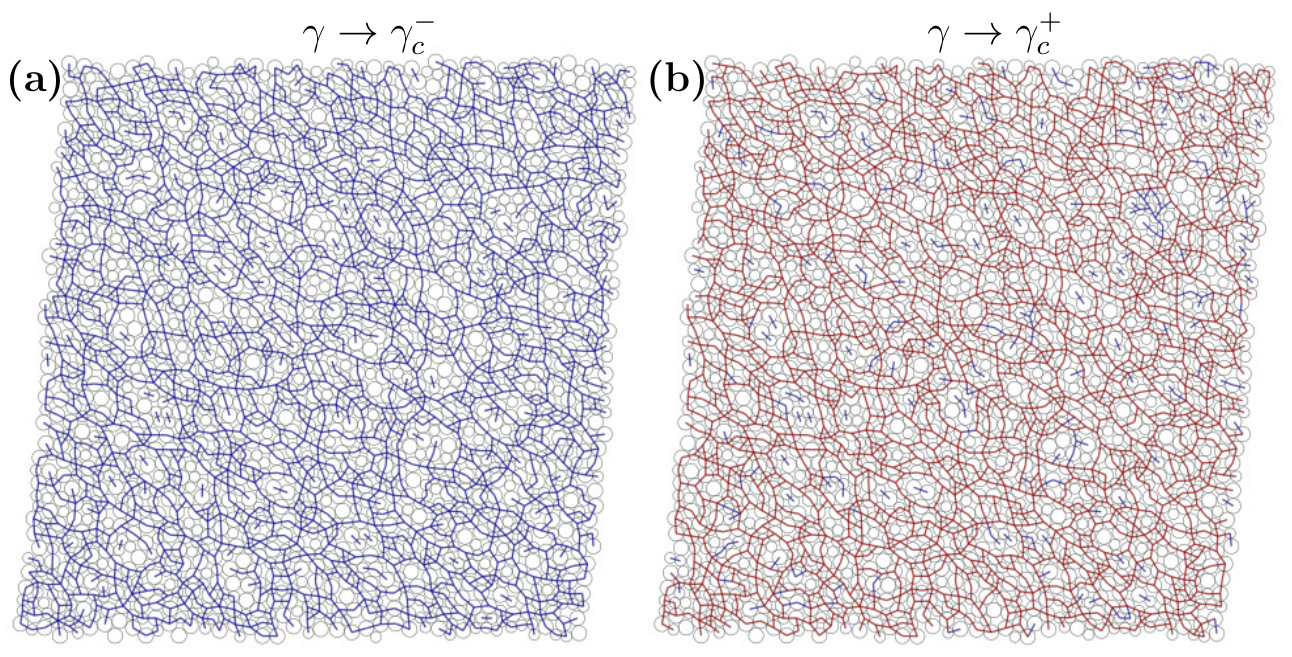}
\caption{Quasi-statically sheared configurations of frictional disks {\bf(a)} just below the shear jamming threshold $\gamma_c$ and {\bf(b)} just above $\gamma_c$. The blue bonds represent geometric contacts between particles and the red bonds represent contacts with finite forces (mechanical contacts). The geometric coordination number $Z_G$ increases smoothly across the transition whereas the mechanical coordination number $Z_M$ shows an abrupt jump from $0$ to $d +1$,  allowing us to precisely locate the transition.} 
\label{geometric_mechanical_contact_network_fig}
\end{figure}

The shear jamming transition of quasi-statically sheared disks has been the subject of several recent studies \cite{bi2011,sarkar2013,vinu2015,baity2017emergent,urbani2017shear,behringer2018physics,otsuki2020shear}. 
Experiments with frictional photoelastic disks \cite{howell1999stress,majmudar2007jamming}, pioneered by Bob Behringer,
show that such a shear induces jamming for a range of densities $\phi_{SJ}$ below the random close packing density $\phi_{RCP}$ (where packings of frictionless disks become rigid) \cite{bi2011}. A theoretical model using solely the stresses near the shear jamming transition showed that a broken translation symmetry indeed emerges in `force-space' \cite{sarkar2013, sarkar2015}. In other computational studies, the shear jamming transition has been sought to be understood in geometrical terms \cite{vinu2015,henkes2016rigid,vinutha2019force,morone2019jamming}.
Whereas indications of the jamming transition as a critical phenomenon have been extensively explored for packings of frictionless  particles \cite{o2003jamming,olsson2007critical,ramola2016disordered,ramola2017scaling}, critical aspects of shear jamming, if present, are largely unexplored. In this paper we report a dynamical hallmark of critical behaviour, namely the divergence of timescales of relaxation, as the shear jamming transition is approached, which we compute by studying the relaxation of forces in the system. We also describe a new procedure of identifying contacts between particles which can be used to precisely locate the shear jamming transition. Although we present results specifically for a system of disks in two dimensions ($d = 2$), we have also performed the same analysis in $d = 3$, and find that similar observations hold for the higher dimensional case ($d = 3$) as well \cite{vinutha_thesis}.

\subsection{Simulating Shear Jamming}
\label{models_methods_section}

\begin{figure}
\includegraphics[scale=0.55,angle=0]{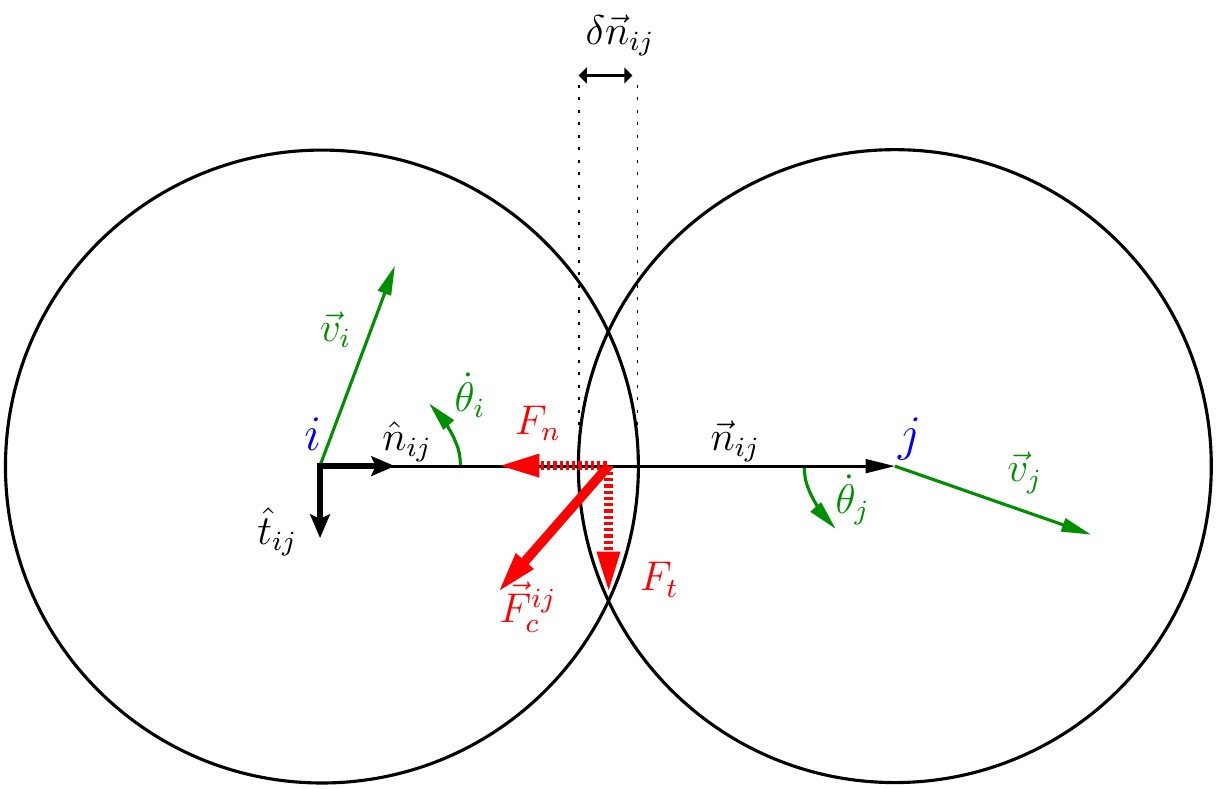}
\caption{A schematic of two particles in contact. The normal force $F^{ij}_n$ and tangential force $F^{ij}_t$ are computed using a linear-spring model (Eq. (\ref{fricpot})).} 
\label{cundall_strack}
\end{figure}

We consider repulsive soft disks with  frictional contact interactions modelled by a Hookean potential (linear-spring dashpot model \cite{cundall}). This model has been used to study the shear jamming transition in several recent studies \cite{vinu2015,vinutha2019force,silbert2010,grob2016}. The normal and tangential components of the contact force $\vec{F}^{ij}_c$ between particles $i$ and $j$ are given by
\begin{equation}
\vec{F}^{ij}_c = \underbrace{(\kappa_n \delta \vec{n}_{ij} - m_{\textmd{eff}} \zeta_n v_n \hat{n}_{ij})}_{\textmd{normal}} -  \underbrace{(\kappa_t \Delta \vec{s}_{t} + m_{\textmd{eff}} \zeta_t v_t \hat{t}_{ij})}_{\textmd{tangential}}. 
\label{fricpot}
\end{equation} 
Here $\kappa_n$ and $\kappa_t$ are elastic constants, while $\zeta_n$ and $\zeta_t$
are damping coefficients for the normal and the tangential velocities respectively. $m_{\textmd{eff}} = 
\sqrt{\frac{m_i m_j}{m_i+m_j}}$ is the effective mass of the two particles in contact. 
$\delta \vec{n}_{ij}$ is the overlap between the spheres in contact along the line joining the centers of the two particles (see Fig. \ref{cundall_strack}).
$\Delta \vec{s}_t$ is the 
tangential displacement vector between the two disks 
from the point of contact. $v_n$ and $v_t$ are the normal and the tangential components of the relative velocity $\vec{v}_i - \vec{v}_j$ between the two 
particles. The maximum value of the tangential force $F^{ij}_t$ is given by the Coulomb criterion 
$F^{ij}_t \le \mu F^{ij}_n$, where $F^{ij}_n$ is the normal force, and $\mu$ is the friction coefficient.
The equation of motion for each particle is given by
\begin{eqnarray}
\nonumber
m_i \frac{d \vec{v}_i}{dt} &=& \sum_j \vec{F}_c^{ij} - \eta \vec{v},\\
I_i \frac{d^2 \theta_i}{dt^2} &=& \sum_j \vec{F}_t^{ij} \frac{\sigma_i}{2} - \eta \frac{d \theta_i}{dt}.
\label{dynamics_eq}
\end{eqnarray}
The summation is over all particles $j$ that are in contact with particle $i$. $I_i$ is the moment of inertia of particle $i$ and $\vec{F}_t^{ij}$ is the tangential force acting on particle $i$ due to particle $j$. Global damping is introduced through the viscous term $ -\eta \vec{v}$, \textmark{which may, {\it e. g.}, be thought of arising from the friction between particles and the bottom plate in two dimensional granular experiments.} As is clear from Eq. (\ref{dynamics_eq}), $m/\eta$ sets the natural unit of time. 
This damping ensures that mechanical equilibrium is achieved within a reasonable computational time \cite{cundall}. 
 Similarly,  we also damp the rotational motion $\dot{\theta}_i$ and $\dot{\theta}_j$ (see Fig. \ref{cundall_strack}) of the particles with the same damping coefficient $\eta$. 
The magnitude of the force at each contact and the total force on each particle are given by
\begin{equation}
|F_c|^{ij}= \sqrt{\left(F^{ij}_n\right)^2 + \left(F^{ij}_t\right)^2}
 ~~\textmd{and} ~~
|F_{\textrm{tot}}|^i = \left|\sum_{j} \vec{F}^{ij}_c\right|.
\label{total_force_equation}
\end{equation}
Here $|~...~|$ represents the magnitude of the vector. 

In our simulations we consider a fixed  system size of $N_{\textmd{total}}=2000$ disks, distributed as a $50:50$ mixture with diameters $\sigma$ and $1.4 \sigma$, with all particles having an equal mass $m_i = m$.
The simulation units we use are: length $[L] = \sigma$, energy $[E] = \kappa_n \sigma^2 /2  \equiv \epsilon$, mass $[M] = m$, and time $[T] =  3 m/\eta =  \sigma \sqrt{m/\epsilon}$.
The input 
parameters in our simulations are $\kappa_n = \kappa_t=2$, $\zeta_n = 3 \sqrt{\epsilon/m}/\sigma$, 
$\zeta_t =\zeta_n/2$, and the friction coefficient $\mu = 1$. The isotropic jamming density for packings of frictionless soft harmonic discs is $\phi_{RCP} \approx 0.843$ \cite{o2003jamming,silbert2010}. 


In order to characterize which configurations reach force balance and which do not, we monitor the following system averaged quantities in our simulations
\begin{eqnarray}
\nonumber
&&\langle |F_c| \rangle = \frac{1}{N_{\textmd{contacts}}} \sum_{ij} |F_c|^{ij},\\
&&\langle  |F_{\textrm{tot}}| \rangle = \frac{1}{N_{\textmd{total}}} \sum_{i} |F_{\textrm{tot}}|^i.
\label{fc_ftot_definition}
\end{eqnarray}
These are the average magnitude of the contact forces and the average magnitude of the force balance in the system respectively. Here $N_{\textmd{contacts}}$ is the total number of contacts between particles that are geometric neighbours, i.e. have a finite overlap or are just touching.

Finally, we use two different protocols to generate force balanced configurations of sheared frictional disks: a Discrete Element Method (DEM) to simulate the relaxation of the system after an affine shear, and an athermal quasi-static shear followed by DEM relaxation (AQS + DEM). 

\subsubsection{Discrete Element Method (DEM)}

 In the DEM protocol, the sequence of configurations are generated as follows {\bf (i)} an affine deformation is applied to the particle coordinates in small increments of $\delta\gamma = 0.01$ and {\bf (ii)} the system is then relaxed using DEM dynamics.  Lees-Edwards boundary conditions are used in this relaxation step. 
The DEM relaxation is similar to well-known molecular dynamics
simulations \cite{cundall}. 
First, we compute the forces acting on each particle using the force-displacement law given in Eq. (\ref{fricpot}) and then, we use these forces to update the positions and velocities 
of the particles.
We terminate the dynamics of the system when a stopping criterion is reached, which in our case (as in \cite{vinutha2019force}) is when the average total force falls below a threshold $\langle |F_{\textmd{tot}}| \rangle \le 5 \times 10^{-12}$.  

Our initial configurations are produced by starting from a hard disk fluid at a low density $\phi =  0.5$. We then apply a fast initial compression using Monte Carlo simulations, i.e. by compressing the box uniformly such that the maximum number of overlaps do not exceed $0.1 \times N_{total}$. Overlaps are removed by running NVT Monte Carlo simulations for the hard-sphere potential, we continue the compression till the desired density is reached. Our data has been averaged over atleast $15$ independent initial samples. These simulations have been performed using the LAMMPS software \cite{lammps}.

\subsubsection{Athermal Quasi-Static (AQS) Shear + DEM}

In order to ensure that our results are independent of protocol we generate force balanced states using an alternate method. This involves two steps:\\ {\bf (i)} The application of a shear deformation (with $\delta\gamma = 5 \times 10^{-5}$) and relaxation using a {\it frictionless} force law (AQS shear). This frictionless force law is given by only the normal component in Eq. (\ref{fricpot}). We then minimize the energy of the system using the conjugate gradient method. 
This generates a sequence of energy minimum configurations at different values of strain. Such a sequence of frictionless sheared configurations has been employed previously  \cite{vinu2015,vinutha2016geometric,vinutha2019force} to analyse geometric aspects of shear jamming. \\
{\bf (ii)} Using each of the energy minimized configurations vs. strain, we apply a uniform compression ($\sigma_{\textmd{tol}} = 5 \times 10^{-5}$) which leads \textmark{overlaps between the geometric contacts of an order equal to $\sigma_{\textmd{tol}}$, which are unbalanced forces. This is one way to apply load on this system.} The system is then allowed to relax using frictional DEM dynamics (which involves the full force law in Eq. (\ref{fricpot})). 
We terminate the dynamics when the average total force falls below the threshold value $\langle |F_{\textmd{tot}}| \rangle \le 5 \times 10^{-12}$.


\subsection{Geometric Contacts versus Mechanical Contacts}

The shear jamming transition is intimately linked with the formation of shear-induced contacts between particles. Indeed, the process can be thought of as a bath of `rattler' particles that contribute increasingly to the contact network as the shear is increased \cite{cates1998jamming}. Conversely, the fraction of non-rattlers serves as a reliable parameter with which to group various global properties of the system near the transition \cite{bi2011}.
In our simulations however, the relaxation dynamics ensure that in the final state there is either a system spanning contact network of force-bearing contacts, or identically zero forces on all contacts. To better understand the structure of the contact network near the shear jamming transition we  quantify the average coordination of the particles in three different ways, we define: {\it 1.} a geometric coordination number $Z_G$ that measures all contacts between particles (force-bearing or otherwise), {\it 2.} a coordination number $Z_{GB}$ defined using a bootstrap procedure where we recursively remove particles that have only one contact from the network, and {\it 3.} a mechanical coordination number $Z_M$ that only counts force-bearing contacts between particles. In all cases, we report the average coordination number for particles which are not rattlers, i.e., those that have more than one contact. 

\begin{figure}
\includegraphics[scale=0.47,angle=0]{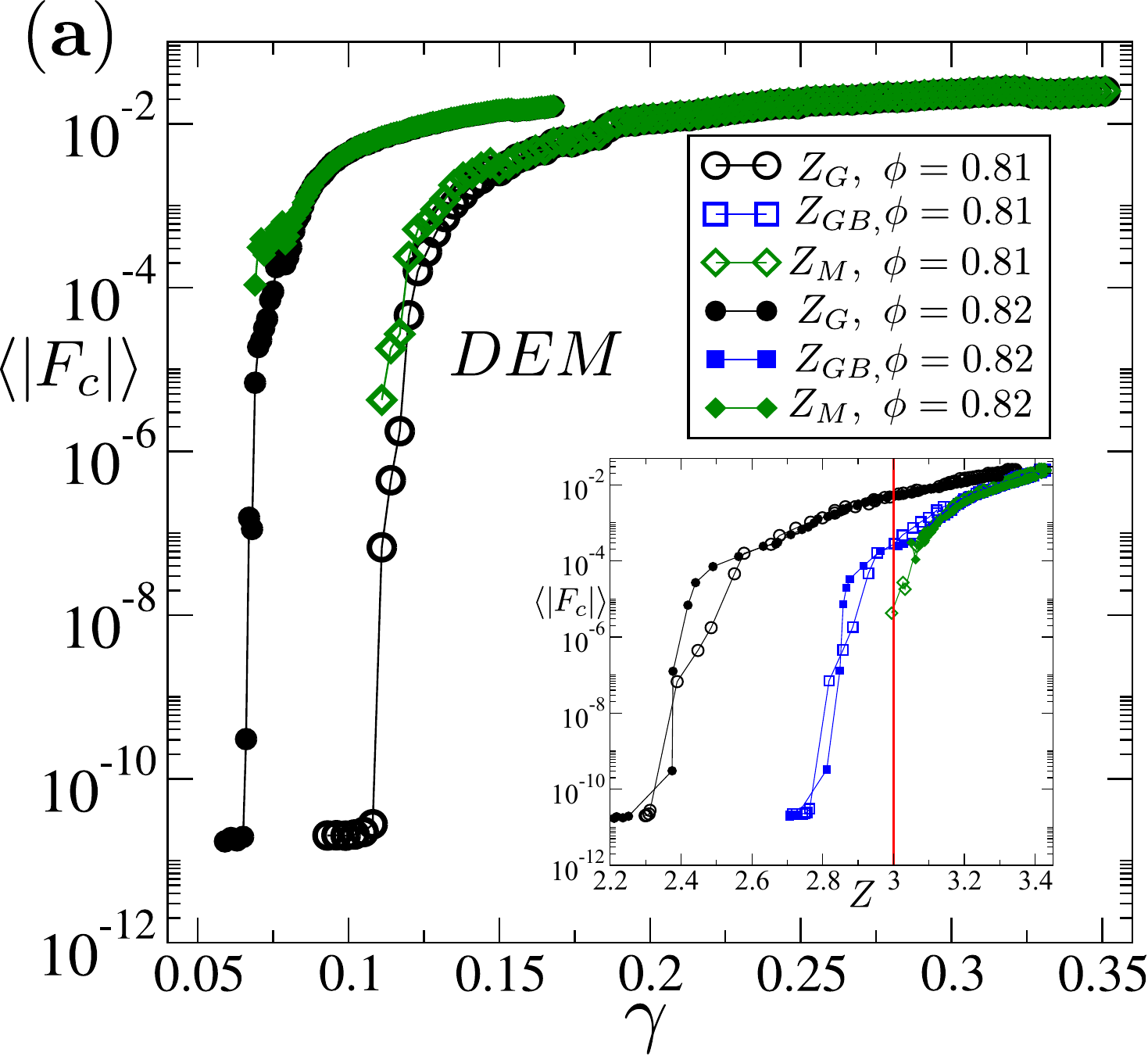}
\includegraphics[scale=1,angle=0]{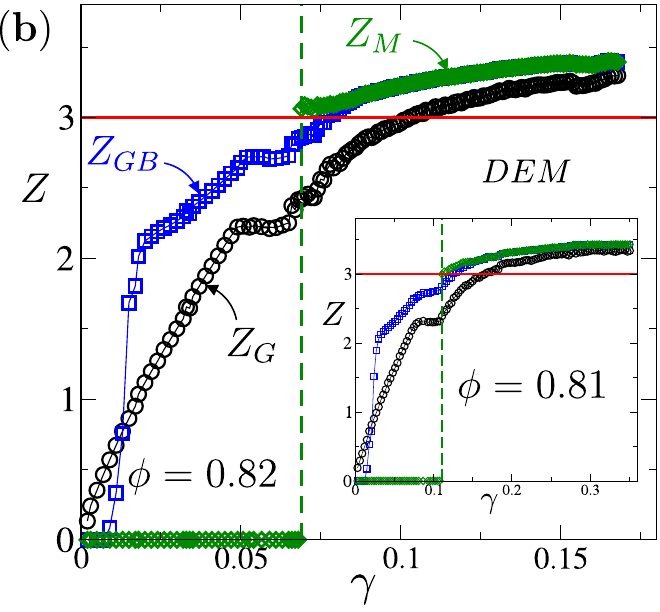}
\caption {
{\bf(a)} The average contact force $\langle |F_c| \rangle$ computed for the geometric and the mechanical contact network as a function of the global strain $\gamma$. {\bf(Inset)} Variation of   $\langle |F_c| \rangle$ as a function of the three different coordination numbers $Z_G, Z_{GB}$ and $Z_M$  using the DEM protocol for two different densities $\phi = 0.81$ and $\phi = 0.82$. The vertical line marks $Z= 3 (=d+1)$. {\bf (b)} The three different coordination numbers $Z_G, Z_{GB}$ and $Z_M$ as a function of strain for density $\phi = 0.82$ and  {\bf (Inset)} $\phi = 0.81$. The horizontal line represents $Z= 3$. We find that the mechanical coordination number $Z_M$ (green points) displays an abrupt jump from $0$ to $\gtrsim d+1$ allowing a precise determination of the transition point. The coordination using the geometric bootstrap procedure $Z_{GB}$ (blue points) is continuous below and closely follows the evolution of $Z_M$ above the transition.}
\label{coordination_figure}
\end{figure}

\subsubsection{Geometric Coordination Number ($Z_G$) }

Two particles are defined to be geometric neighbours if the distance $r_{ij}$ between them obeys $r_{ij} \leq \sigma_{ij}$, i.e. if they have a finite overlap, or are just touching (here $\sigma_{ij} = \sigma_i + \sigma_j$ is the summation of their radii). \textmark{The geometric coordination number $Z_G$ is then defined to be
\begin{equation}
Z_G = \frac{N_{\textmd{geometric contacts}}}{N_{\textmd{total}}}.
\label{geometric_contact_definition}
\end{equation}
$N_{\textmd{geometric contacts}}$ is the sum of the number of geometric contacts of $N_{\textmd{total}}$ particles.}
Geometric contacts can have precisely zero force (within numerical error in our simulations). This occurs because once two particles are separated by a distance $r_{ij} = \sigma_{ij}$, there is no force between them and therefore no relative evolution. Indeed, below the shear jamming transition even though there are a finite number of contacts in the system, we find that {\it all} contacts have either zero force, or forces comparable in magnitude to the total force on each particle. There is therefore no separation in scale between the force balance and the contact forces.
The average contact force computed for the geometric contact network is shown in  the inset of Fig. \ref{coordination_figure} {\bf (a)}, showing a smooth increase with increasing $Z_G$. 

\subsubsection{Coordination using `Geometric Bootstrap' ($Z_{GB}$)}
\textmark{Particles with less than two geometric contacts are defined to be `geometric rattlers'. The definition of rattlers in this case is motivated by the fact that particles with only a single contact can never be in force balance as long as there are finite forces in the system, whereas particles with two or more neighbors can.}
The definition of rattlers as particles with less than or equal to one contact is ambiguous for the following reason. Let us suppose a geometric rattler with a single contact is removed from the system. This then leads to a decrease in the number of contacts of its neighbour, which can in-turn become a geometric rattler itself. If one were to follow this procedure to completion, this would require a recursive removal of rattler contacts from the system until all particles have either $0$ or $\ge 2$ contacts. We term such a procedure `Geometric Bootstrap' (GB), and the coordination of the system after such a recursion is denoted by the symbol $Z_{GB}$. 
\textmark{The geometric boostrap coordination number $Z_{GB}$ is then defined to be
\begin{equation}
Z_{GB} = \frac{N_{\textmd{geometric contacts}}}{N_{\textmd{total}} - N_{\textmd{geometric rattlers}}}.
\label{geometric_contact_definition}
\end{equation}
$N_{\textmd{geometric contacts}}$ is the sum of the number of geometric contacts of non-rattler particles, {\it i.e.,}$N_{\textmd{total}}- N_{\textmd{geometric rattlers}}$.}
The average contact force computed for the contact network after the geometric bootstrap procedure is shown in the inset of Fig. \ref{coordination_figure} {\bf (a)}. This displays a smooth increase with $Z_{GB}$, qualitatively similar to the increase with respect to $Z_G$. 
However, crucially this {\it increases} the average coordination of the system (see Fig. \ref{coordination_figure} {\bf (b)}). The condition $Z_{GB} = d +1$ therefore occurs at a lower strain value as compared to $Z_{G}$. By analyzing the force balance of the system using force bearing contacts, we find that $Z_{GB}$ predicts the shear jamming transition with surprising accuracy. The geometric bootstrap procedure therefore represents a purely geometric way to precisely locate the transition. We expect that the precision with which an isostatic value of $Z_{GB}$ identifies the shear jamming transition will get better at larger friction values and become exact in the infinite friction limit, based on earlier analyses \cite{vinutha2019force}.

\begin{figure}
\includegraphics[scale=0.45,angle=0]{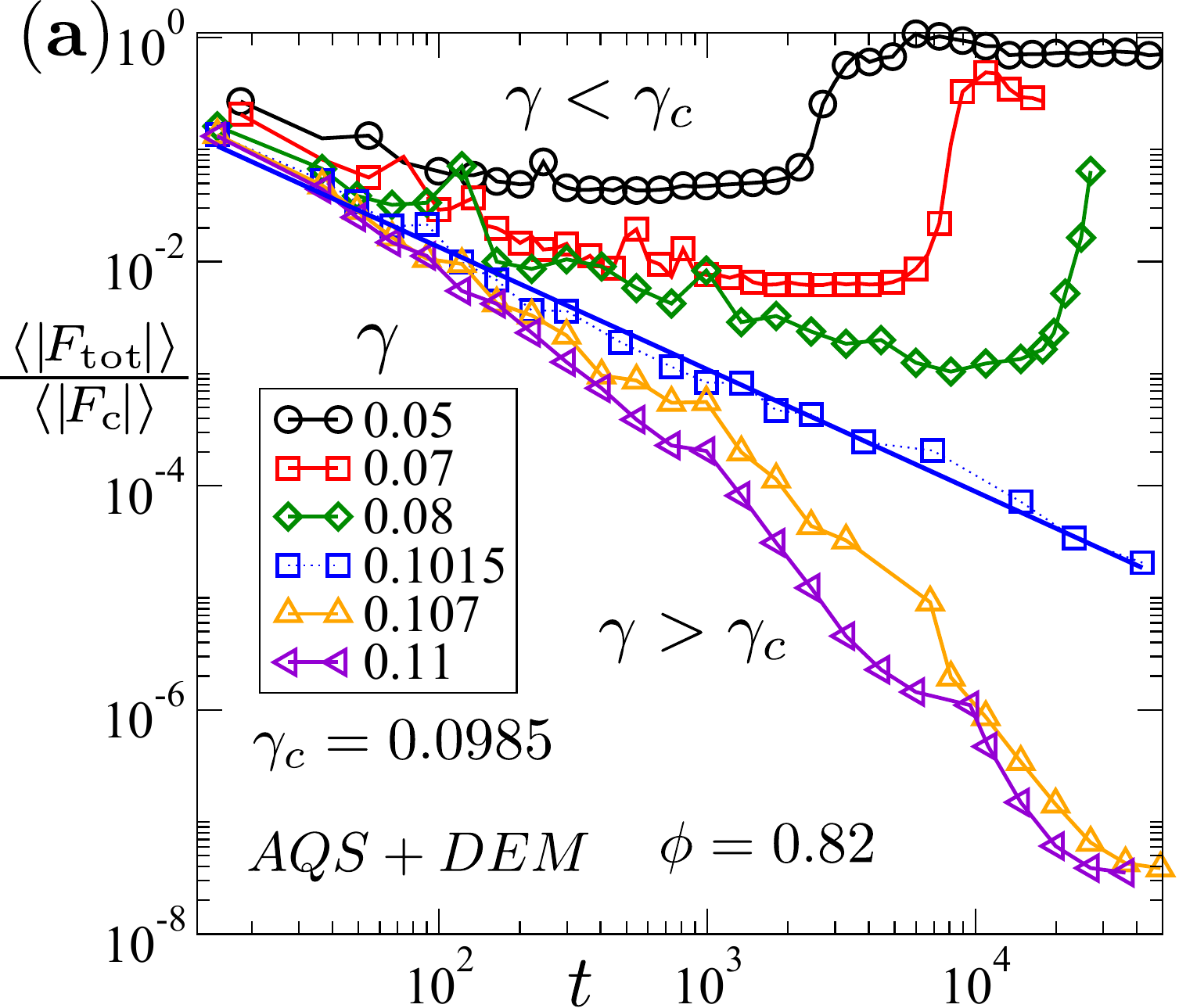}
\vspace{0.5cm}
\includegraphics[scale=0.45,angle=0]{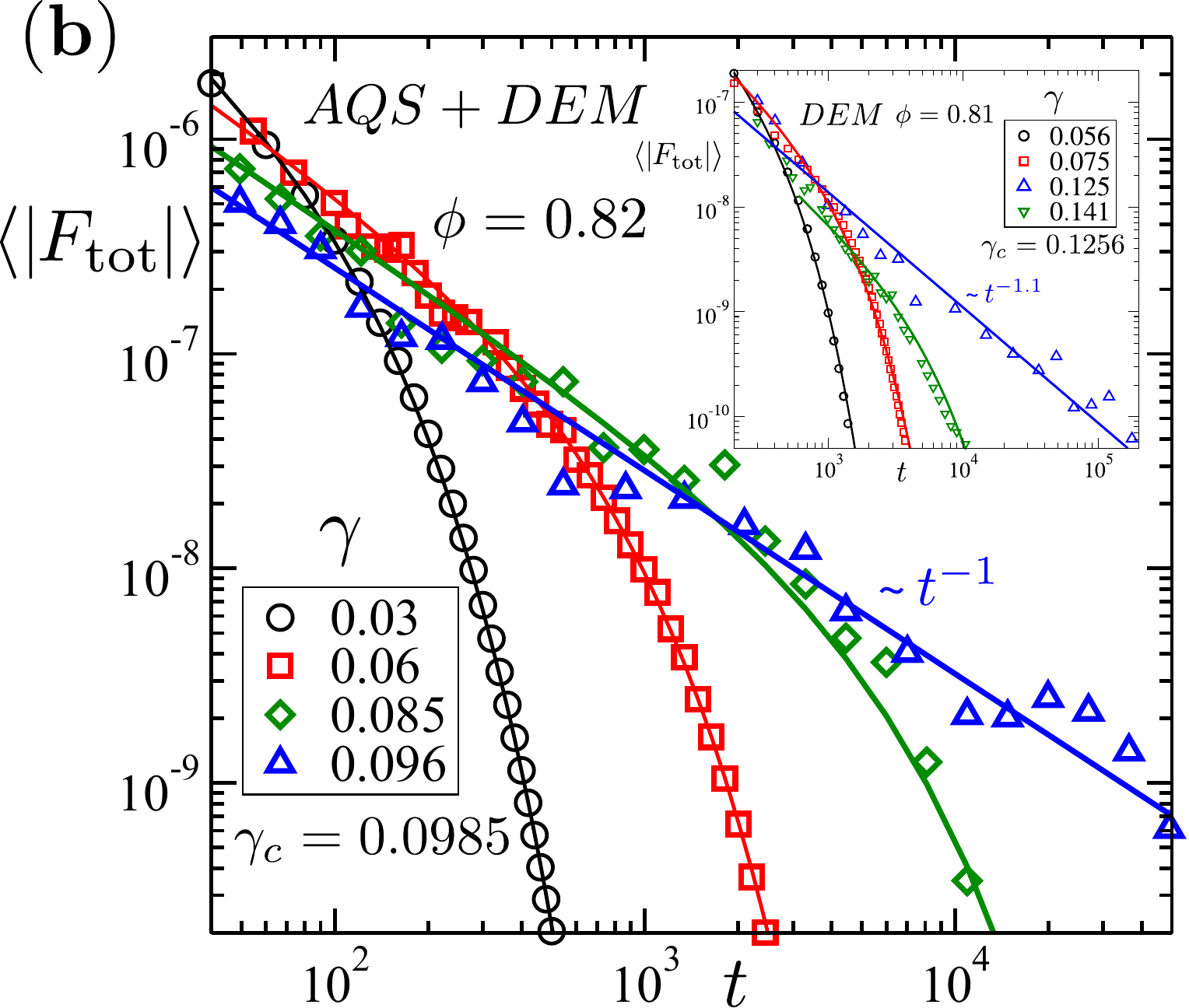}
\caption{
{\bf (a)}  The evolution of the ratio $\langle |F_{\textrm{tot}}| \rangle/ \langle |F_c| \rangle$ for different values of the global strain. The data represents a single run using an initial configuration at a starting density $\phi = 0.82$ and subsequently evolution using the AQS+DEM protocol.
{\bf (b)} The relaxation of the average total force on the particles $\langle |F_{\textmd{tot}}| \rangle $ as a function of time at different strain values using the AQS+DEM protocol from a starting density $\phi = 0.82$. We estimate $\tau^{*}$ (which we plot in Fig. \ref{timescale_figure_2} {\bf (a)}) by fitting the decay to a form $\langle |F_{\textmd{tot}}| \rangle (t) \propto \exp(-t/\tau^*) t^{-\alpha}$.  At the critical point the relaxation of $\langle F_{\textmd{tot}} \rangle $ is fit well with a power law $t^{-\alpha}$ with  $\alpha \approx 1$. {\bf(Inset)} The relaxation of the average total force $\langle |F_{\textmd{tot}}| \rangle $ using the DEM protocol from a starting density $\phi = 0.81$. The critical relaxation is fit well with a power law $t^{-\alpha}$ with  $\alpha \approx 1.1$.
}
\label{timescale_figure_1} 
\end{figure}

 \subsubsection{Mechanical Coordination Number ($Z_M$)}
  
Finally, we define a mechanical coordination number $Z_M$ which only accounts for contacts that carry finite forces between particles which are in {\it force balance}.  A contact  on a particle is defined to be force bearing if the magnitude of the contact force is much larger than the total force on the particle. We choose this criterion to be $|F_c|^{ij}/|F_{\textrm{tot}}|^{i} > 10^2 $, where $|F_{\textrm{tot}}|^{i}$
is the total force on particle $i$ defined in Eq. (\ref{total_force_equation}). The value $10^2$ is chosen so that in our simulations,  numerical errors will not lead to a violation of Newton's third law. Specifically, if particle $j$ is counted as a mechanical contact of particle $i$, particle $i$ should in turn be counted as a mechanical contact of $j$. We find $10^2$ to be the smallest threshold that guarantees this feature.
Particles with less than two force-bearing contacts are defined to be `mechanical rattlers'. The mechanical coordination number $Z_M$ is then defined to be


\begin{equation}
Z_M = \frac{N_{\textmd{mechanical contacts}}}{N_{\textmd{total}} - N_{\textmd{mechanical rattlers}}}.
\end{equation} 
\textmark{$N_{\textmd{mechanical contacts}}$ is the sum of the number of mechanical contacts of force-bearing particles.}
In our simulations, we only compute $Z_M$ for force balanced configurations, which we define to be configurations with an average contact force $\langle |F_c| \rangle / \langle |F_{\textrm{tot}}| \rangle  > 10^{3}$.
The average contact force $\langle |F_c| \rangle$ computed using this mechanical contact network is shown in the inset of Fig. \ref{coordination_figure} {\bf (a)}.  $\langle |F_c| \rangle$ displays a discontinuity at $Z_M = 3$, allowing us to precisely locate the shear jamming transition $\gamma_c$. 
Remarkably, this value of $\gamma_c$ is very close to the value obtained from the condition $Z_{GB} = d +1$   (see Fig. \ref{coordination_figure} {\bf (b)}).
Note that in Fig. \ref{coordination_figure} we have shown results for two different densities, with the transition occurring at a lower strain value  $\gamma_c$ as the packing fraction is increased. We find that the that precise value of the transition point depends on the starting density, and that this dependence is the only significant influence of the density, consistent with previous studies \cite{bi2011,sarkar2015,vinutha2019force}.

\subsection{Diverging Timescale}

We next study the relaxation dynamics of the system close to the shear jamming transition $\gamma_c$, as identified by the mechanical coordination number in the previous section.
After each strain step, we relax the system using the DEM method and we monitor the total force as a function of time. The DEM dynamics that we use ensures that the average total force $\langle |F_{\textmd{tot}}| \rangle$ on the particles tends to zero at late times. However, the configurations below and above the shear jamming transition display fundamentally different relaxation behaviour.

 \subsubsection{Critical Relaxation}

Above shear jamming $\gamma > \gamma_c$, at late times a finite force bearing network of contacts emerges.  The magnitude of the force on each contact is well separated from the magnitude of the force balance on the corresponding particle  $ |F_c|^{ij} / |F_{\textmd{tot}}|^{i}  \gg 10^2$. Below shear jamming $\gamma < \gamma_c$, although the mechanical coordination remains precisely zero, an increasing number of geometric contacts develop as the shear is increased. 
These states are characterized by inadequate force balance on each grain $| F_c|^{ij} / |F_{\textmd{tot}}|^{i} < 10^2$. 

Starting with a single initial configuration we measure the evolution of $\langle |F_c| \rangle$, $\langle |F_{\textrm{tot}}| \rangle$ (defined in Eq. \ref{fc_ftot_definition}), along with the ratio $\langle |F_{\textrm{tot}}| \rangle/ \langle |F_c| \rangle$.
 In Fig. \ref{timescale_figure_1}{\bf (a)} we plot this ratio $\langle |F_{\textrm{tot}}| \rangle/ \langle |F_c| \rangle$ for a single run using the AQS+DEM protocol with a starting density $\phi = 0.82$.
 In the $\gamma > \gamma_c$ regime, as the system relaxes, the average total force on each disk $\langle |F_{\textmd{tot}}| \rangle$ evolves to zero, while $\langle |F_c|\rangle$ remains finite. The ratio therefore decays to zero at large times.
In contrast, for $\gamma < \gamma_c$, as the system relaxes $\langle |F_{\textmd{tot}}| \rangle$ and $\langle |F_c|\rangle$ together decay to zero. We find that after an initial decay this ratio begins to increase.
When both $\langle |F_{\textmd{tot}}| \rangle$ and   $\langle |F_c|\rangle$  are within numerical error in our simulations, the ratio $\langle |F_{\textmd{tot}}| \rangle / \langle |F_c| \rangle $ saturates to a constant value of $\mathcal{O}(1)$. At the critical point $\gamma_c$ we find that the relaxation of  $\langle |F_{\textrm{tot}}| \rangle/ \langle |F_c| \rangle$ is well-fit with a power law $t^{-\alpha}$. These lines in Fig. \ref{timescale_figure_1} {\bf (a)} and {\bf (b)} mark a separatrix between the two regimes of decay, above and below the shear jamming transition.
We find from our simulations that $\alpha \approx 1$ for both densities and protocols that we have studied.

 \subsubsection{Estimating a Time Scale ($~\tau^*$)}
 
\begin{figure}
\includegraphics[scale=0.45,angle=0]{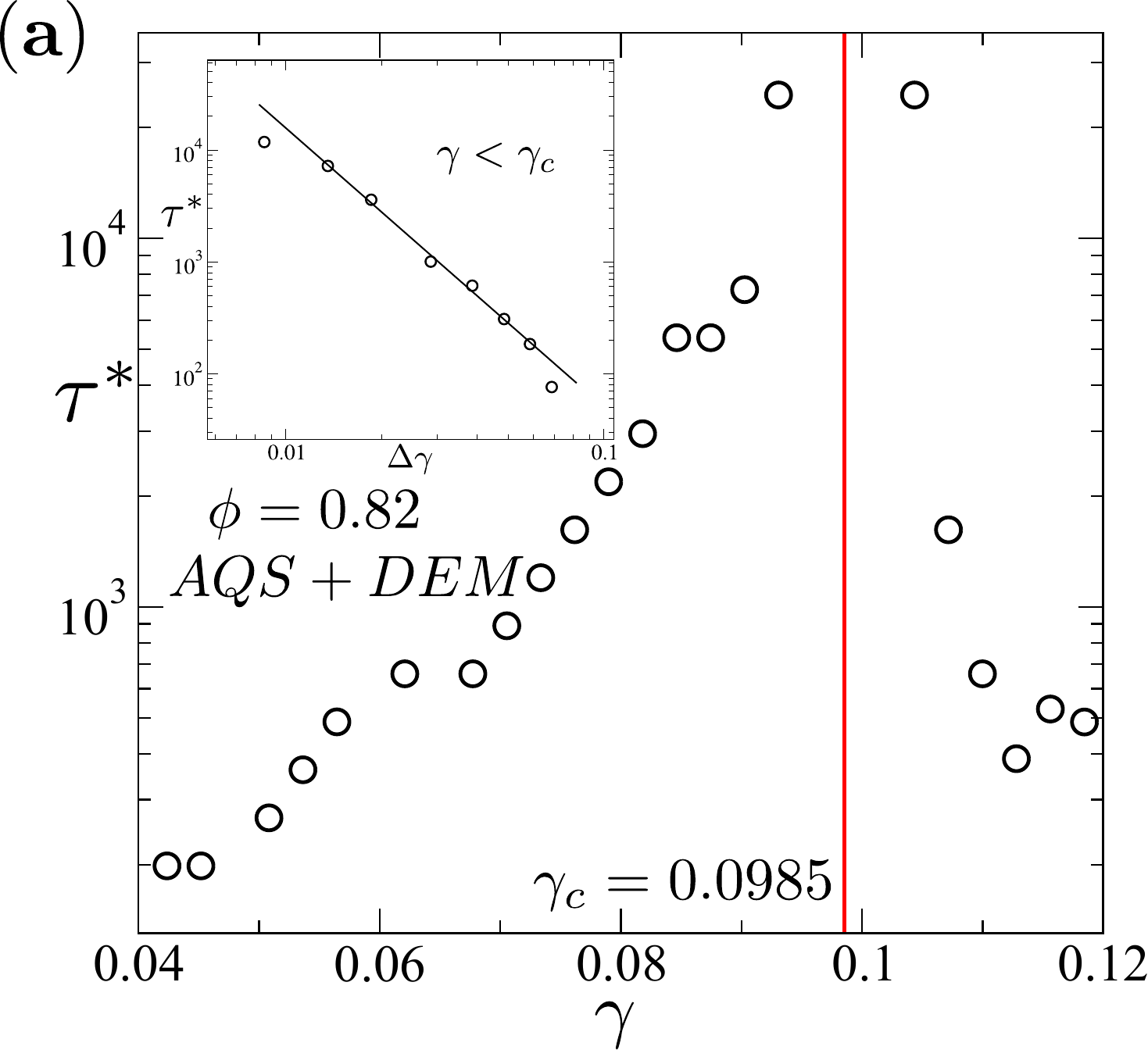}
\hspace{-10mm}
\includegraphics[scale=0.45,angle=0]{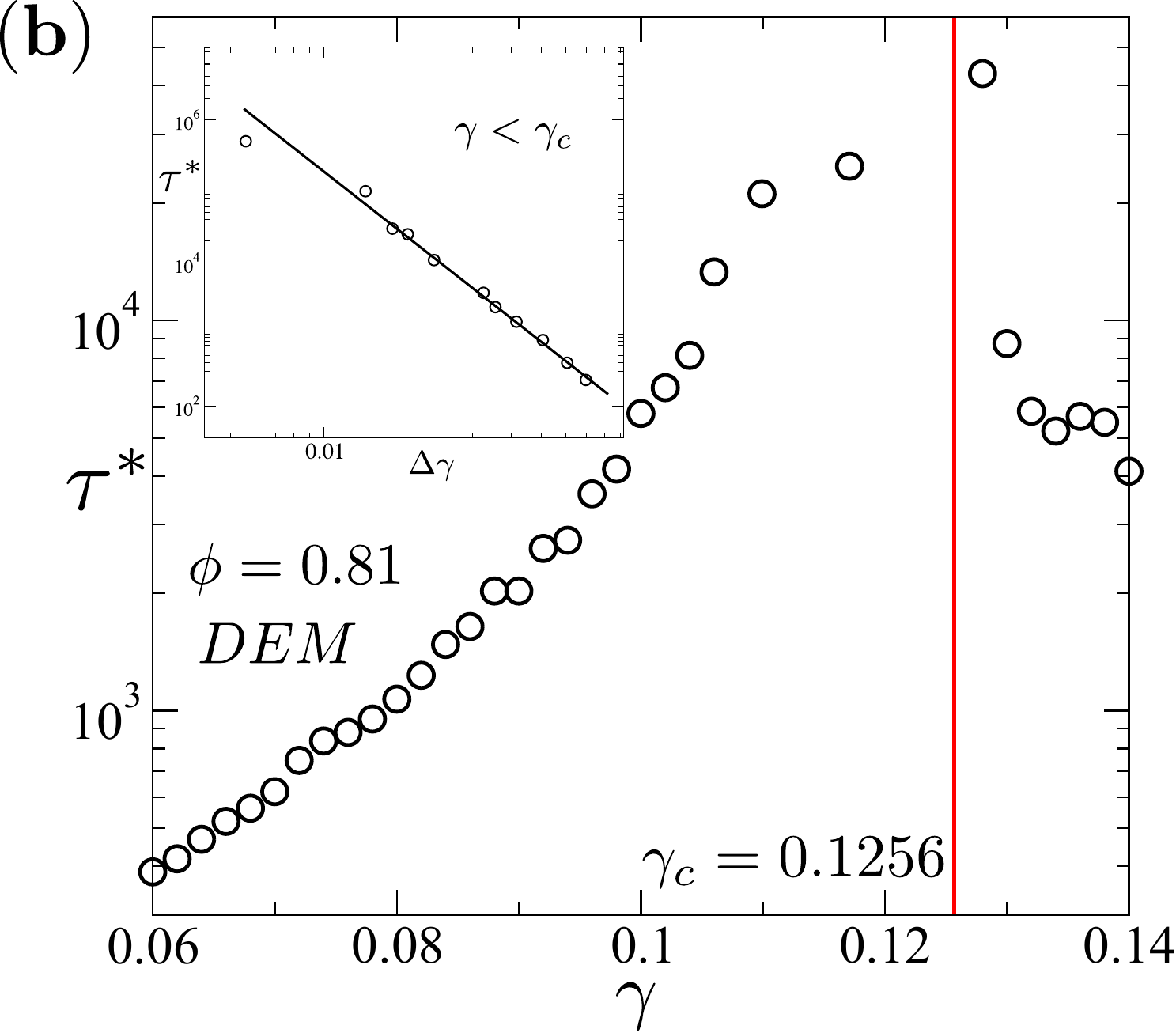}
\caption{{\bf (a)} Plot of $\tau^{*}$, the extracted time scale as a function of the strain
using the AQS+DEM protocol from a starting density $\phi = 0.82$.  {\bf (Inset)} The same data plotted in log-log scale. {\bf (b)} Plot of $\tau^{*}$ using the DEM protocol from a starting density $\phi = 0.81$. As the shear jamming transition is approached, this time scale becomes larger, displaying a diverges at the critical strain value $\gamma_c$. {\bf (Inset)} The same data plotted in log-log scale. The divergence can be reliably fit with a power law at large $\Delta \gamma$. }
\label{timescale_figure_2}
\end{figure}

 We next estimate a characteristic time scale $\tau^{*}$ for the relaxation of the forces in the system at each strain value. We extract this time scale from the relaxation of the average total force on particles $\langle | F_{\text{tot}}| \rangle$ as a function of time. Since we follow the evolution of a single initial configuration for this measure, this decay in our simulations has a large amount of noise. Therefore at each strain value, we average the data in log bins to obtain a smooth curve. 
  In Fig. \ref{timescale_figure_1} {\bf (b)}, we plot the relaxation of $\langle |F_{\textmd{tot}}| \rangle $ as a function of time $t$ at different strain values close to the shear jamming transition.  We find that these decays are exponential both above and below the transition, which we identify using the discontinuity in the mechanical coordination number $Z_M$ (see Fig. \ref{coordination_figure} {\bf (b)}).
We find that the decay of $\langle |F_{\textmd{tot}}| \rangle$ can be fit well with a decay law of the form $\langle |F_{\textmd{tot}}| \rangle (t) \propto \exp(-t/\tau^*) t^{-\alpha}$ with  $\alpha \approx 1.1$, which can be used to obtain an estimate for $\tau^*$. In practice, we estimate $\tau^{*}$ by first removing the power-law component of the decay and fitting the remaining data to a purely exponential decay.

In Fig. \ref{timescale_figure_2} we plot the extracted timescales $\tau^*$ for two cases  (a) a single run using the AQS+DEM protocol at a density $\phi = 0.82$ and (b) a single run using the DEM protocol at a density $\phi = 0.81$.
As the shear jamming transition is approached from above and below, the time scale in both cases becomes larger and displays a divergence at the transition. At the critical point the relaxation is fit well with a power law $t^{-\alpha}$ with  $\alpha \approx 1.1$.  
 In the inset of Fig. \ref{timescale_figure_2} we plot the extracted timescales $\tau^*$ as a function of the distance to the transition $\Delta \gamma$. We find that the approach is well characterized by a power law $\tau^* \propto \Delta \gamma^{-\nu}$, with $\nu \approx 3.4$ for the DEM case and  $\nu \approx 2.5$ for AQS+DEM.

The large variation in the values of the exponents $\alpha$ and $\nu$ arises from the 
difficulties associated with attaining force balance and consequently estimating timescales close to the transition.  In this paper, we have established that  the shear jamming transition displays dynamical hallmarks of critical behaviour. In order to ascertain the dependence of these exponents on protocol and density, a more detailed study over a larger range of densities near the shear jamming transition is required.

Finally, based on the data presented in Fig. \ref{timescale_figure_2}, we conclude that for a given microscopic evolution rate of the system (which corresponds to the evolution dynamics we use in our simulations), as the global shear $\gamma$ approaches the critical value $\gamma_c$ from below, the time taken to relax to a state where all the inter-particle contacts have uniformly zero force, diverges.  
Above $\gamma > \gamma_{c}$, the system settles into a state characterized by finite forces between particles, with the timescale also increasing as $\gamma \to \gamma_{c}^{+}$.

\section*{Discussion}
\label{discussion_section}

In this paper we have simulated amorphous packings of frictional disks subject to an external shear close to the shear jamming transition. We found that the transition can be characterized by an abrupt jump in the number of force bearing contacts between particles, $Z_M$. This transition is accompanied by a diverging timescale $\tau^*$ that characterizes the time required by the system to attain force balance.

However, several open questions remain. The origin of this timescale divergence near the transition remains to be elucidated. Recent studies have found a viscosity divergence and dynamical slowing down at the density-driven frictionless jamming transition \cite{peter2018crossover,ikeda2020universal}. In that case an isolated low energy mode in the Hessian accompanies the slowdown. The divergence of timescales we find is driven by shear strain rather than density.
We have analyzed the Hessians associated with the contact networks that form in our system, however we have not been able to identify such a mode. It would be interesting to understand the origin of this difference between the two types of jamming transitions. \textmark{A thorough investigation of the effect of density on the shear-jamming strain, over a wider range of densities than considered here, will be carried out in the future, following work reported in \cite{das2019unified}}.

Finally, in experiments of sheared frictional particles below $\phi_{SJ}$, the stresses are mediated essentially along one-dimensional structures known as `force chains'. Such states are termed `fragile' \cite{bi2011,seto2019shear}, and are force balanced configurations with $Z_G < d + 1$. We have not been able to capture the fragile states in our simulations since the forces in the system are then under-constrained, and finding a solution to particle-level force balance therefore necessitates the emergence of additional constraints involving the positions of the particles. This situation seems to be rare in our simulations. Thus, further investigations along the lines reported here are needed to elucidate the properties of fragile states.

\vspace{5mm}
\begin{acknowledgments}
We dedicate this work to the memory of Bob Behringer whose insightful experiments with granular matter are an inspiration to us all. We acknowledge support from the Indo-US Science and Technology Forum (grant no. IUSSTF/JC-026/2016). The work of BC and KR has been supported by NSF-CBET 1605428.  BC acknowledges a fellowship from the Simons Foundation. SS and HAV gratefully acknowledge TUE-CMS, SSL, JNCASR, for computational resources and support. SS acknowledges support through the J C Bose Fellowship, DST, India.
\end{acknowledgments}

 
\bibliography{2D_Shear_Jamming_Bibliography}
\bibliographystyle{apsrev4-1}
\end{document}